\documentstyle[aps,prd]{revtex} 

\draft

\begin{document}
%@twocolumnfalse\endcsname
\title{%
\hbox to\hsize{\normalsize\rm August 1999
\hfil MPI-Pth/99-34}
\vskip 36pt
 On the  Relation Between Quantum Mechanical and Classical Parallel
Transport
}
\author {J. Anandan}
\address{Department of Physics and Astronomy, University of South
Carolina, Columbia, South Carolina 29208}

\author{L.~Stodolsky}
\address{Max-Planck-Institut f\"ur Physik 
(Werner-Heisenberg-Institut),
F\"ohringer Ring 6, 80805 M\"unchen, Germany}
%\date{August 13, 1999}
\maketitle
\begin{abstract} We explain how the kind of ``parallel transport''
of a 
wavefunction used in discussing the Berry or Geometrical phase
induces
the conventional parallel transport of certain real vectors.
These real vectors are associated with operators whose commutators
yield diagonal
operators; or in Lie algebras those operators whose commutators are
 in the (diagonal) Cartan subalgebra.
\end{abstract}
%\pacs{PACS numbers: }

\vskip2.0pc

In discussing the Berry~\cite{berry} or Geometrical phase one uses
the concept
of 
a ``parallel transport" of a quantum mechanical state $\psi$, by
which is meant
\begin{equation}\label{para}
 < \psi(t)\vert\dot \psi(t)>= 0 
\end{equation}
One may also consider  a complete (orthonormal)  basis set of
states  $\vert n>$
obeying this condition. If the time-dependent  
   states $\vert n(t)>$ are   obtained from
a basis of   initial states $\vert n >$ by a unitary transformation
$U$ (as
would be generated by a hamiltonian)
\begin{equation}\label{upar}
 U(t)\vert n >=\vert n(t)> 
\end{equation}
we can say we have an orthonormal  ``frame" undergoing this kind of
parallel
transport~\cite{us}. 
In this case the ``parallel transport" condition, namely
\begin{equation}\label{paraa}
 < n(t)\vert\dot n(t)>= 0
\end{equation}
for all $n$, can also be written as
\begin{equation}\label{udot}
<n(t)\vert \dot U U^{\dagger}\vert n(t)>=0 
\end{equation}
  (Quantities without a time argument 
refer to the fixed basis, while those with an argument (t) refer to
the moving basis, thus $\vert n>= \vert n(0)>$.)
Now the ``parallel transport" and ``moving frames" implied by these
equations
are  not  the same as those of usual differential geometry. Rather,
there, in the viewpoint where one studies a euclidean frame moving
in a higher dimensional space and then restricts
 to a submanifold~\cite{flanders}, there is 
 a set of  real vectors ${\bf e}_a$
 instead of quantum mechanical state
vectors,
 and parallel transport among a set of vectors 
$a,b,c...$ on the submanifold means
\begin{equation}\label{rea} 
 {\bf {\dot e}}_a(t)\cdot{\bf e}_b(t)=0
\end{equation} 
for all pairs $a,b,c...$~ in the submanifold. That is, the set
${\bf e}_a$ are not
a complete set, but rather form a moving subspace in a larger
space. In this formulation the dot symbol means the ordinary
derivative in the
ambient space, while in the ``intrinsic''  formulation of
differential geometry
 the dot symbol would mean the covariant
derivative with a connection.  

 This condition~Eq~[\ref{rea}], which we might call ``classical"
parallel
transport, looks
quite different from  Eq~[\ref{paraa}].
What  is the relationship between the two kinds of
``parallel
transport'', if any?

 It seems there should be some such relationship. For example, in
our treatment~\cite{us},
 of  the
geometric phase in SU(2), where $U$ is an SU(2) group
element, we could 
view the ``quantum mechanical parallel transport'' as inducing
the ``classical
parallel transport''
of the $x$ and $y$ vectors of a ``dreibein" sliding, but not
rotating, on the sphere.

Here we would like to briefly elucidate why this is and to indicate
how  to generalize the idea, including its application to higher
groups.
  Briefly, we will show how  the
condition Eq~[\ref{paraa}] for a complex basis leads to a
``classical'' parallel transport, Eq~[\ref{rea}], of certain
vectors 
associated with the problem, such as the $x$, $y$ vectors of the
``dreibein". 

 Our first task is to identify the vectors ${\bf e}_a$,
which we do as follows. 
 Consider a complete set of operators or matrices
$\lambda_a$, like the generators of a Lie group, complete in the
sense that they transform among each other under $U$.  That is,
there are the time dependent operators
$\lambda (t)=U(t)\lambda_a U^{\dagger}(t)$, which may be 
reexpressed in
terms of the original, fixed, $\lambda$. These then 
generate the  vectors ${\bf e}_a$ via (summation convention)
\begin{equation}\label{lat}
\lambda_a(t)= U(t)\lambda_a U^{\dagger}(t)={\bf
e}_a^j(t)\lambda^j
\end{equation}

 If we choose the $\lambda_a$ such that $Tr(\lambda_a \lambda_b)=N
\delta_{ab}$, where $N$ is a normalization factor, we
can write explicitly
\begin{equation}\label{comp}
{\bf e}_a(t)^j =1/N~Tr[\lambda_a(t)\lambda^j]={\cal   
T}r[\lambda_a(t)\lambda^j]
\end{equation}
where we define ${\cal T}r$ to include the normalization factor. 
Furthermore, with hermitian $\lambda_a$ the ${\bf e}_a(t)$ are
real.
The scalar product of two vectors is then given by the trace
of the product of the corresponding $\lambda$, as in ${\bf 
e}_a(t)\cdot{\bf e}_b(t)= {\cal T}r[\lambda_a(t)\lambda_b(t)]$. 
The definition of the $\lambda(t)$ is chosen so that
 $< n(t)\vert\lambda(t)\vert m(t)>=< n\vert\lambda\vert m>$. 

Now a main  point of ~\cite{us} was that the information conveyed
by
the condition Eq~[\ref{paraa}] or
Eq~[\ref{udot}]
could be interpreted, in the group theoretical context, by saying
that in the ``local frame''
there was no rotation with respect to the subspace of
diagonal generators, that is in the  Cartan subspace.
 We can formulate this point in a general manner
  by viewing the evolution of the states as being determined by a
hamiltonian $h(t)$, where  $h(t)=i\dot U
U^{\dagger}$. (We reserve the symbol $H$ for the more usual
hamiltonian, which
however is absent in the present considerations. 
 $H$ includes the 
``dynamical phase'' which usually~\cite{berry,us} has been removed
from the problem
before we get to Eq [\ref{para}]).   Thus
\begin{equation}\label{h}
i\vert\dot n(t)>= h(t)\vert n(t)> 
\end{equation}
and Eq~[\ref{udot}] states that the diagonal elements of $h$ are
zero in the moving basis:
\begin{equation}\label{hd}
  <n(t)\vert h(t)\vert n(t)> =0 
\end{equation}
  
We would now like to explain how Eq~[\ref{hd}] can lead to
$  {\bf {\dot e}}_a(t)\cdot{\bf e}_b(t)=0$ among some of the ${\bf
e}(t)$.
 The desired quantity ${\bf {\dot e}}_a(t)\cdot{\bf e}_b(t)$
may   be found from
\begin{equation}\label{tr}
{\bf {\dot e}}_a(t)\cdot{\bf e}_b(t)={\cal T}r[\dot\lambda_a(t)
\lambda_b(t)]=i{\cal T}r[~[h(t),\lambda_a(t)]\lambda_b(t)]
\end{equation}
where  we use the  equation of motion $i\dot
\lambda_a(t)=[h(t),\lambda_a(t)]$ following from the definition of 
 $\lambda_a(t)$.
 Rearranging the last expression we have
\begin{equation}\label{trr}
{\bf {\dot e}}_a(t)\cdot{\bf
e}_b(t)=i{\cal T}r[h(t)[\lambda_a(t),\lambda_b(t)]~]
\end{equation}

  Now consider  two $\lambda$'s such that their commutator gives
a diagonal matrix, $[\lambda_a,\lambda_b]= (~diag~)$. Applying $U$,
the same holds in the moving basis for the $\lambda_a(t)$
namely
\begin{equation}\label{diag}
<m(t)\vert[\lambda_a(t),\lambda_b(t)]\vert n(t)>\sim
\delta_{nm}
\end{equation}

 For such a pair  in Eq~[\ref{trr}], while  $h$ has no diagonal
elements,
the commutator has   only diagonal elements. But this  gives zero
for
the trace, and hence ${\bf {\dot e}}_a(t)\cdot{\bf
e}_b(t)=0$.

 We thus arrive at our conclusion:
 ``classical parallel transport'', Eq~[\ref{rea}], 
  follows as a result of  ``quantum parallel transport''
Eq~[\ref{paraa}]
for those vectors ${\bf e}_a(t)$ whose corresponding    commutators
among the $\lambda_a$ yield  diagonal matrices. In an abbreviated
language with a ``matrix valued vector'' $e_a^j\lambda^j$, we
could say ``for those vectors whose mutual commutators are
diagonal''.

 In group
theory this is the requirement that the commutator lie in the
Cartan subalgebra, when the latter, as usual, has been chosen
diagonal.  
Precisely this was the case in our SU(2) example~\cite{us} where 
the $S_x=\lambda_x$ and
$S_y=\lambda_y$ generators are
the two non-Cartan
generators. Their  commutator yields only the 
Cartan generator $S_z=\lambda_z$, which in the usual choice of
basis is
diagonal. This is why under a $U$ obeying Eq~[\ref{udot}]
they undergo parallel transport in the sense of Eq~[\ref{rea}].
Note  however, that it is necessary to explicitly take  the Cartan
operators  diagonal.

{\centerline {\bf Comments}}

 A striking difference between the
quantum
Eq~[\ref{paraa}] and the classical Eq~[\ref{rea}] is that  in the
classical
case the concept is linear;  if two vectors are parallel
transported then
their sum is also. However, for the parallel transport of Eq~
[\ref{para}], as may be easily verified, this is not true.

This implies that Eq~[\ref{paraa}] is not in general
preserved under linear transformation; a new ``frame"  $\vert
n'>=\sum
u_{n'n}\vert n >$ will not in general be parallel transported, even
if the $\vert
n >$ are. Thus a full statement of the problem involves a
specification as to which set of vectors satisfy Eq~[\ref{paraa}],
as reflected in the necessity to
 choose a definite basis,  one in which the Cartan generators are
diagonal.
 This helps in clarifying the following potential misunderstanding:
We might be
tempted to conclude  that when dealing with real quantities, as
with
orthogonal rotations, that Eq~[\ref{paraa}] follows  simply
from unitarity and thus represents no further information. That is,
 given that all quantities are
real,
$< n(t)\vert\dot n(t)> + < \dot n(t)\vert n(t)> = 2 < n(t)\vert\dot
n(t)>=0$ follows simply from preservation of the norm.
Thus we would be lead by our above arguments to the nonsensical
result that any orthogonal transformation will
automatically induce parallel transport.
However, this argument would neglect the requirement that the
Cartan operators  be diagonal. In fact for real
orthogonal representations the generators are antisymmetric, or in
the above notation, the $\lambda$ are pure imaginary. But  
we need the Cartan operators in diagonal form, and
antisymmetric operators cannot be brought to diagonal form without
introducing a complex basis. Thus complex numbers are reintroduced
and Eq~[\ref{paraa}] does indeed represent a second condition, and
not  just simply unitarity or
preservation of the norm.

We would like to thank the Institute of Advanced Studies,
Jerusalem, for its hospitality in the spring of 1998, when this
work was begun.

\end{document}